\documentclass[twocolumn,aps,prl,superscriptaddress,amsmath,amssymb,floatfix,footinbib]{revtex4-2}

\usepackage[colorlinks=true,citecolor=blue,linkcolor=magenta]{hyperref}
\usepackage[utf8]{inputenc}
\usepackage[english]{babel}
\usepackage[T1]{fontenc}
\usepackage{amsmath,amsfonts,amssymb}
\usepackage{bm}            
\usepackage{siunitx}
\usepackage[version=4]{mhchem}
\usepackage{graphicx}
\graphicspath{{./Figures/}}
\makeatletter
\long\def\nat@makecaption#1#2{%
  \vskip\abovecaptionskip
  \begingroup
    \small
    \parindent\z@ \leftskip\z@ \rightskip\z@ \parfillskip\z@\@plus1fil\relax
    \noindent{\bfseries #1\nobreakspace\textbar}\nobreakspace #2\par
  \endgroup
  \vskip\belowcaptionskip}
\AtBeginDocument{%
  \let\@makecaption\nat@makecaption
  \renewcommand\fnum@figure{Fig.\nobreakspace\thefigure}%
}
\makeatother

\newcommand{\WEL}{\affiliation{Laboratory of Wave Engineering, \'Ecole Polytechnique F\'ed\'erale de Lausanne (EPFL), Switzerland}}
\newcommand{\HW}
 {\affiliation{Huawei Technologies Sweden AB, R\&D Centre}}

\begin{document}
\raggedbottom
\title{Beating the nonreciprocal isolation limit of integrated circulators by Floquet leakage interference \\}
\author{Zhe Zhang}
\altaffiliation{Current address: Department of Electrical Engineering, Ginzton Lab, Stanford University, USA}
\WEL

\author{Haoye Qin}
\altaffiliation{Current address: Tsinghua Shenzhen International Graduate School, Tsinghua University, Shenzhen 518055, China}
\WEL

\author{Junda Wang}\WEL
\author{Qiaolu Chen}\WEL
\author{Zhechen Zhang}\WEL
\author{Alireza Mafi}\WEL
\author{Ahsan Altaf} \HW
\author{Romain Fleury}
\email{romain.fleury@epfl.ch}
\WEL

\begin{abstract}
Time-modulation using semiconductor switches is a promising route to magnetless integrated
nonreciprocal devices, as they can offer large modulation depths and high speeds. Yet, chip-scale devices are capped by the finite off-state capacitance of the switches, which induces detrimental leakage of the input wave to the isolated port,
imposing a ceiling on the nonreciprocal isolation. Such leakage has largely constrained the development of nonreciprocal integrated systems at high frequency. Here, we beat this limit by using Floquet interference between leakages. In a time-Floquet switched-resonator circulator, two coherent leakages reach the isolated port: the release of the stored wave at the resonator's ring-down frequency, and the direct input leakage at the carrier frequency. We show that it is possible to create conditions under which the two
leakages destructively interfere, and even completely cancel, yielding perfect isolation despite operating with non-ideal semiconductor switches. We experimentally confirm leakage interference in a 65-nm CMOS microwave Floquet circulator, reaching 40-dB isolation, which is more than 20 dB higher than the natural switch isolation. We also demonstrate that leakage interference has inherently fast dynamics, establishing itself within a single modulation period, allowing us to reverse the circulation chirality in 0.6~ns. Our results pave the way toward high-frequency integrated chips with ultra-high nonreciprocal isolation.
\end{abstract}
\maketitle

\begin{figure*}[t]
  \centering
  \includegraphics[width=\textwidth]{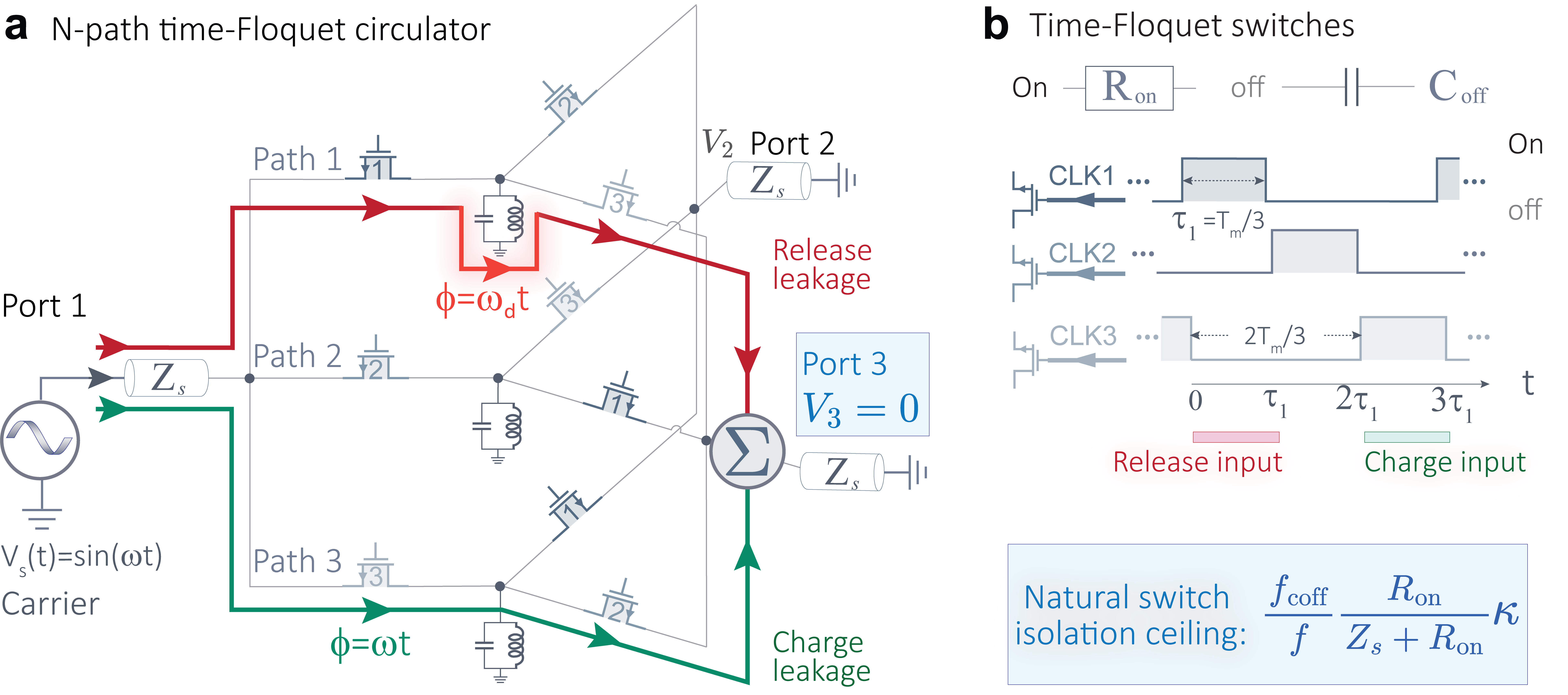}
  \caption{\textbf{Coherent cancellation of leakages in a time-Floquet circulator.} We consider a time-Floquet circulator based on an $N$-path switched-resonator network (\textbf{a}). The switches are periodically modulated at period $T_m$ by three cyclically shifted control clocks (\textbf{b}). Each switch is driven by its gate voltage, toggling between the equivalent circuits $R_{\mathrm{on}}$ (on) and $C_{\mathrm{off}}$ (off). For a wave input at port~1, an ideal clockwise circulator delivers it only to port~2, with port~3 nominally isolated ($V_3=0$). Yet two leakages inevitably reach port~3: the intended ring-down of the resonant reservoir (release leakage, red), and a parasitic path through the off-state capacitance while the input wave charges the resonator (charge leakage, green). This off-state capacitance caps the isolation at the natural switch isolation, fixed by the switch rather than the circuit (bottom, \textbf{b}). We beat this limit by creating conditions under which the two leakages interfere destructively.}
  \label{fig:concept}
\end{figure*}

\emph{Nonreciprocity} is the key physical attribute of electronic devices such as circulators, which enable full-duplex communication by isolating the transmitter and receiver modules in antenna front ends, where more than $20$\,dB of isolation is typically demanded\cite{calozElectromagneticNonreciprocity2018,sounasNonreciprocalPhotonicsBased2017,naguluNonMagneticNonReciprocalMicrowave2021,naguluDoublingWirelessCapacity2024,reiskarimianAnalysisDesignCommutationBased2018}. Integrated microwave circulators realize nonreciprocity by modulating a circuit parameter periodically in time, such as a conductance or a capacitance, and are therefore referred to as time-Floquet circulators\cite{yuCompleteOpticalIsolation2009,fangPhotonicAharonovBohmEffect2012,fangRealizingEffectiveMagnetic2012c,sounasGiantNonreciprocitySubwavelength2013,estepMagneticfreeNonreciprocityIsolation2014,fleurySoundIsolationGiant2014,yuanBlochOscillationUnidirectional2016,reiskarimianMagneticfreeNonreciprocityBased2016,naguluNonreciprocalElectronicsBased2020,sohnElectricallyDrivenOptical2021,herrmannMirrorSymmetricOnchip2022,naguluChipscaleFloquetTopological2022}.
The simplest and most-efficient modulation component in integrated platforms is the transistor
switch, operated as a binary conductance, as it delivers deep, fast
modulation up to its cutoff
frequency $f_{\mathrm{coff}}$, which can reach the terahertz range\cite{dincMillimeterWaveNonMagneticPassive2017,samizadehnikooNanoplasmaenabledPicosecondSwitches2020,samizadehnikooElectronicMetadevicesTerahertz2023}. Yet, the switch is also where the signal leaks:
in its off state it is not an open circuit but a finite capacitance, through which
the input wave reaches the very port meant to be isolated\cite{dincMillimeterWaveNonMagneticPassive2017,tymchenkoQuasielectrostaticWavePropagation2019}. This leakage, intrinsic to the switch rather than the circuit, caps the attainable isolation at a ceiling $\propto f_{\mathrm{coff}}/{f}$, scaled by the small voltage fraction dropped across the switch, as we show below\cite{naguluNonMagneticNonReciprocalMicrowave2021,yang85110GHzCMOSMagneticFree2019}. We term this bound the natural switch isolation: it is severe and tightens as the carrier frequency $f$ rises. Even a 500 GHz-cutoff switch, carefully designed, yields only about 20 dB of isolation at a few-gigahertz carrier, short of what typical front ends demand.\\

An opportunity to beat this ceiling lies in modifying what holds the input wave between
modulation events. A switched capacitor, the standard choice\cite{franksAlternativeApproachRealization1960,ghaffariTunableHighQNPath2011}, is quasistatic: once
the input source is disconnected, its stored state is the voltage, which decays exponentially without accumulating phase. The leakage contributions from different signal paths therefore all carry the carrier-frequency phase and can only add, leaving the ceiling in place. In contrast, a switched resonator keeps ringing throughout the storage at its own ring-down frequency, which can differ from the carrier frequency\cite{notomiWavelengthConversionDynamic2006,galiffiPhotonicsTimevaryingMedia2022}. This frequency offset builds a phase difference between the leakage paths, making their cancellation possible. At the same time, resonances allow for modulation frequencies much lower than the carrier frequency, instead of the fast Nyquist rate that a switched capacitor demands. \\

In this article, we propose and demonstrate the concept of time-Floquet leakage interference, which removes the isolation ceiling and enables large nonreciprocal isolation in modulated integrated systems, despite the intrinsic limitation of semiconductor switches. We theoretically identify two coherent
leakages at the isolated port and show that, albeit distinct in origin and advancing in phase at different frequencies, they can synchronize to perfectly annihilate each other.  We unveil the destructive interference condition and show that points of perfect cancellation necessarily emerge on the carrier frequency--modulation period plane. We then observe the perfect cancellation in a 65-nm CMOS microwave circulator, which exhibits a nonreciprocal isolation of $40$\,dB, far beyond the natural switch isolation of $10$--$20$\,dB.  We also demonstrate the inherently fast dynamics of the interference by reversing the wave circulation chirality in sub-nanosecond time. The time-Floquet leakage cancellation thus imposes no carrier limit of its own: the natural switch isolation drops as the carrier rises, while the leakage cancellation involves only the frequency detuning of the carrier from the resonance.

\begin{figure*}[t]
  \centering
  \includegraphics[width=\textwidth]{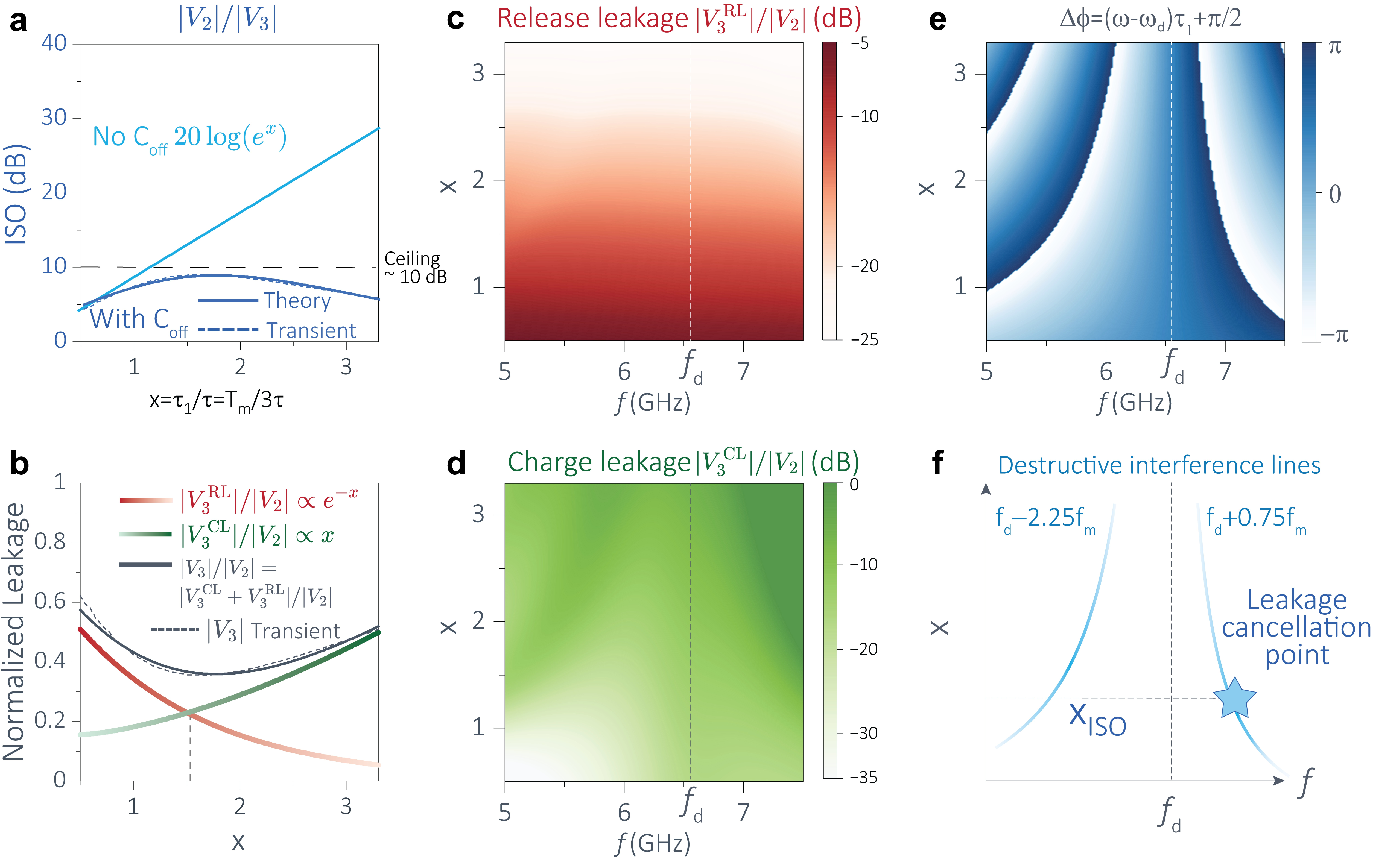}
  \caption{\textbf{Leakage decomposition of the natural switch isolation ceiling and the necessary emergence of a leakage cancellation point.} \textbf{a}, The off-state capacitance caps the isolation at the natural
  switch isolation. For an ideal switch (no $C_{\mathrm{off}}$, light blue) at the ring-down frequency $f_d=6.57$\,GHz, the isolation rises exponentially with the normalized modulation period $x$. For a practical switch of finite cutoff frequency (with $C_{\mathrm{off}}$, deep blue), it saturates at the
  $\sim$10-dB ceiling. \textbf{b}, The ceiling originates from the charge leakage: the release leakage $|V_3^{\mathrm{RL}}|/|V_2|$ (red) decays exponentially with $x$, while the charge leakage $|V_3^{\mathrm{CL}}|/|V_2|$ (green) grows as $1/\kappa(x)$, approximately linearly over the displayed range. Without time-Floquet interference, the total leakage $|V_3|/|V_2|$ can never be made small, and its minimum sets the ceiling (theory lines, with the transient simulation dashed). \textbf{c,d}, The two leakages occupy complementary regions of the $(f,x)$ plane: $|V_3^{\mathrm{RL}}|/|V_2|$ (\textbf{c}) dominates at small $x$, $|V_3^{\mathrm{CL}}|/|V_2|$ (\textbf{d}) at large $x$. Their $x$-dependence is set in the time domain, so the crossover
  exists at every carrier frequency. \textbf{e}, The leakage phase difference of Eq.~(\ref{eq:dphi}) spans the full $\pm\pi$ across the plane, reaching anti-phase along a pair of destructive-interference lines rendered asymmetric about $f_d$ by the capacitive quarter cycle. We focus on the upper line, $f=f_d+0.75f_m$, as it lies closest to $f_d$. \textbf{f}, Schematic of the necessarily emerging cancellation point, where anti-phase meets equal amplitude. On the line, the falling and rising amplitudes match at $x_{\mathrm{ISO}}$ (star), where the two anti-phase
  leakages cancel.}
  \label{fig:ceiling}
\end{figure*}

\section*{Two coherent leakages of time-Floquet circulators}

We consider a three-port time-Floquet circulator built on an $N$-path switched-resonator network with $N=3$ (Fig.~\ref{fig:concept}a). Each switch, implemented as a transistor in integrated platforms, is operated as a binary conductance, with an on-resistance $R_{\mathrm{on}}$ and an off-state capacitance $C_{\mathrm{off}}$, and is driven by one of three control clocks. These clocks are cyclically shifted by $\tau_1=T_m/3$, with $T_m$ the modulation period, in a clockwise sequence, so that exactly one path couples each port at any instant and the wave circulation inherits the same chirality (Fig.~\ref{fig:concept}b). Owing to the cyclic symmetry, a single source at port~1, $V_s(t)=\sin(\omega t)$ with carrier frequency $f=\omega/2\pi$, suffices to characterize the wave circulation, with ports~2 and~3 being the transmitted and the isolated outputs. The nonreciprocity is then quantified by the isolation ratio
\begin{equation}
\mathrm{ISO}\equiv|V_2|/|V_3|,
\label{eq:iso}
\end{equation}
with $V_2$ and $V_3$ the complex phasors of the port voltages at the carrier
frequency $f$. Chiral routing requires storing the input wave between its charge, with the path connected to the source at port~1, and its release, with the source disconnected and the load at port~2 connected, since an instantaneous charge-and-release path would be reciprocal. Each path therefore charges a reservoir, and the resonator is the decisive choice for it: throughout the release, it keeps ringing at its damped ring-down frequency $\omega_d$ (see Methods), which can differ from the carrier frequency, providing a second phase reference on which the leakage cancellation is built. 
 
For an ideal clockwise circulator the input at port~1 reaches only port~2, leaving port~3 isolated ($V_3=0$). In practice, leakage to port~3 arises through two distinct mechanisms (Fig.~\ref{fig:concept}a). The first is the release leakage (RL): after the stored energy is released to port~2, the
residue keeps ringing at $\omega_d$ and escapes to port~3 in the following on-state window (red arrow, Fig.~\ref{fig:concept}a). The second is the charge leakage (CL), named after the charging window: while the input charges the next resonator, the off-state capacitance on its far side leaks part of the wave directly to port~3 (green arrow, Fig.~\ref{fig:concept}a). 
The charge leakage is intrinsic to the switch: independent of how the
reservoir is built, it caps the isolation at the natural switch isolation
\begin{equation}
\mathrm{ISO}_{\mathrm{CL}}\approx
\frac{f_{\mathrm{coff}}}{f}\,
\frac{R_{\mathrm{on}}}{Z_s+R_{\mathrm{on}}}\kappa,
\label{eq:NS}
\end{equation}
with $\kappa$ an order-unity kernel comparing the useful release with the direct charge leakage, defined in Eq.~(\ref{eq:kappa}) of the Methods. This limit is shared by integrated switch-based Floquet circulators\cite{reiskarimianMagneticfreeNonreciprocityBased2016,dincMillimeterWaveNonMagneticPassive2017,naguluNonMagneticCMOSSwitchedTransmissionLine2019},
whose isolation relies on the off state blocking the signal path (boxed
relation, Fig.~\ref{fig:concept}b). The two leakages, however, oscillate at
two different frequencies, the ring-down at $\omega_d$ and the carrier at
$\omega$, so a phase difference accumulates during the storage. Once driven to anti-phase with the amplitudes matched, the two leakages cancel and the isolated-port wave vanishes, beating the limit set by the switch.

\begin{figure*}[t]
  \centering
  \includegraphics[width=0.6\textwidth]{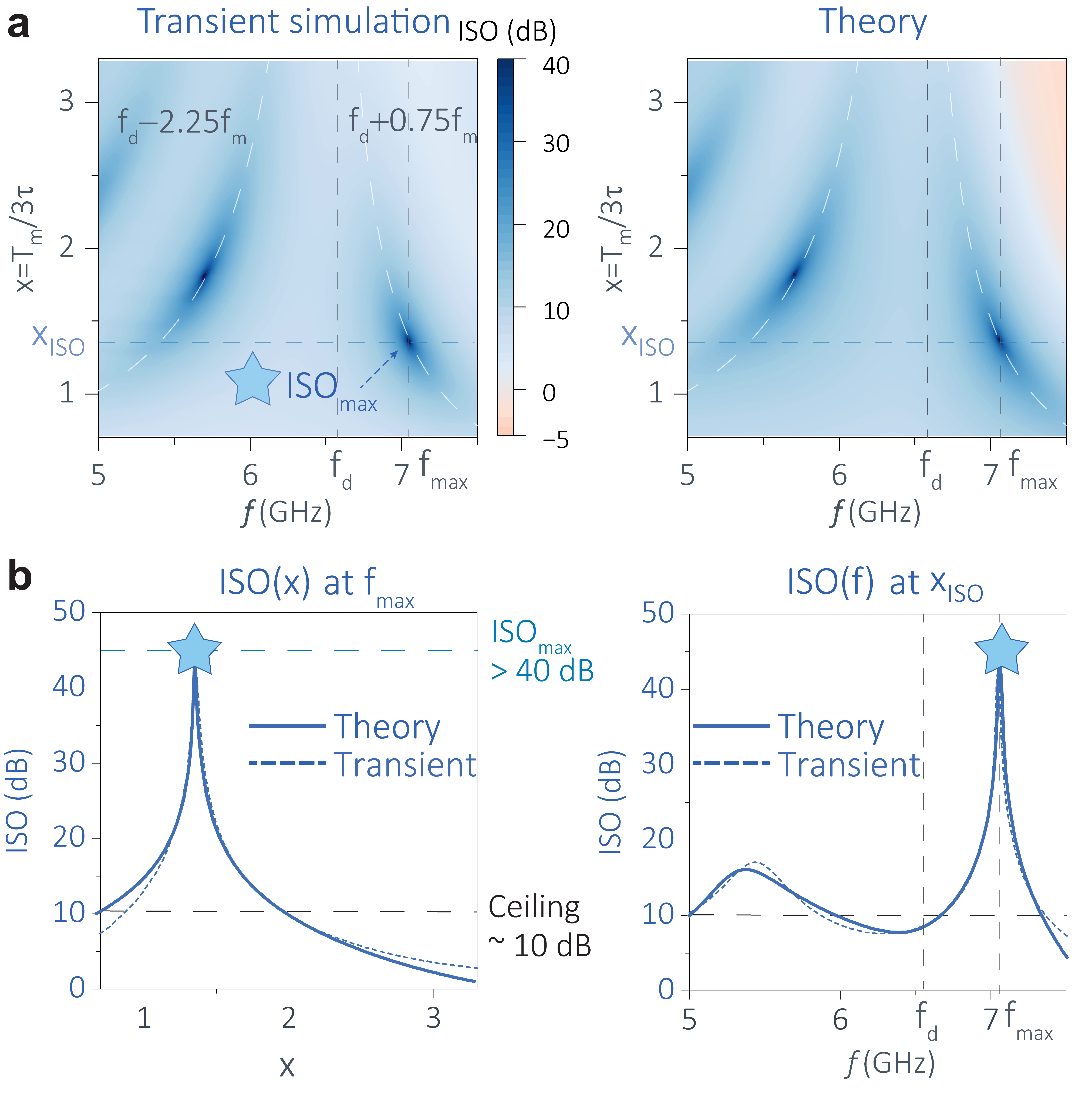}
  \caption{\textbf{Ultra-high isolation from time-Floquet leakage
  interference.} \textbf{a}, The cancellation concentrates along the destructive-interference lines and is perfect at a single point. Isolation over the $(f,x)$ plane, from transient simulation (left) and theory (right). High isolation traces the asymmetric line pair $f=f_d+0.75f_m$ and $f=f_d-2.25f_m$ (white dashed) of
  Fig.~\ref{fig:ceiling}e. The perfect cancellation, $\mathrm{ISO_{max}}$ (star), occurs where a line crosses the equal-amplitude point $x_{\mathrm{ISO}}$ of Fig.~\ref{fig:ceiling}f, defining the corresponding carrier frequency $f_{\mathrm{max}}$. \textbf{b}, At the cancellation point, the isolation rises far beyond the natural switch isolation. Isolation versus $x$ at $f_{\mathrm{max}}$ (left) and versus $f$ at $x_{\mathrm{ISO}}$ (right): at the cancellation point the isolation peaks above $40$\,dB, approximately $30$\,dB beyond the $\sim$10-dB ceiling (grey dashed, for reference).}
  \label{fig:peak}
\end{figure*}

Such cancellation, however, is available only to a reservoir of higher-order dynamics. A first-order capacitor is quasistatic: it holds the amplitude of the wave but freezes its phase, so it must operate as a sampler at the Nyquist rate, $f_m\gtrsim 2f/N$, with $f_m=1/T_m$ the modulation
frequency\cite{tymchenkoQuasielectrostaticWavePropagation2019,naguluUltraWidebandSwitchedCapacitorDelays2021}. The storage is then pinned to the carrier scale, and the anti-phase accumulation cannot be accommodated (see Methods). In contrast, a second-order resonator preserves the carrier's phase over many ring-down cycles, relaxing the modulation frequency to $f_m\sim 2f/(NQ_{\mathrm{load}})$, with $Q_{\mathrm{load}}$ the loaded quality factor. The low modulation frequency provides a long window, and the phase difference, accumulating at the small detuning $\omega-\omega_d$, is able to reach anti-phase within that single window of $\tau_1$.

\section*{Time-Floquet leakage interference}

We numerically study a Floquet circulator based on an ideal switch, without $C_{\mathrm{off}}$, and one with a practical integrated switch with $f_{\mathrm{coff}}=1.23$\,THz, around the ring-down frequency $f_d=6.57$\,GHz (see circuit simulation details in Methods). The relevant timescale of the reservoir is its time constant $\tau$, and the switch open time $\tau_1$ is naturally normalized to it, $x\equiv\tau_1/\tau$.  For the ideal switch, the isolation grows
exponentially with $x$ (Fig.~\ref{fig:ceiling}a). Yet, for the practical switch, it saturates just below $10$\,dB, capped by the natural switch isolation of Eq.~(\ref{eq:NS}). Circuit optimization such as tuning $Z_s$ modifies only the prefactor, not the frequency scaling\cite{dincMillimeterWaveNonMagneticPassive2017,kordBroadbandCyclicSymmetricMagnetless2018}, and the matched-port ($Z_s=50\,\Omega$) isolation of practical integrated circulators accordingly clusters at $10$--$20$\,dB\cite{reiskarimianMagneticfreeNonreciprocityBased2016,dincMillimeterWaveNonMagneticPassive2017,naguluNonMagneticCMOSSwitchedTransmissionLine2019}, the natural switch isolation. Theory and transient circuit simulation agree throughout (Fig.~\ref{fig:ceiling}a).

The leakage behind this $10$-dB ceiling is the coherent sum of two contributions, which only the theory resolves individually. The transmitted voltage amplitude at port 2 follows the charge--discharge envelope of the resonator,
\begin{equation}
|V_2|\propto 1-e^{-x},
\label{eq:v2}
\end{equation}
and varies slowly when $x>1$, so the isolation of Eq.~(\ref{eq:iso}) is governed by the port-3 leakage. The latter is the coherent sum of the
two leakage contributions,
\begin{equation}
\begin{aligned}
\mathrm{ISO}^{-1}&=\frac{|V_3^{\mathrm{RL}}+V_3^{\mathrm{CL}}|}{|V_2|},\\
\frac{|V_3^{\mathrm{RL}}|}{|V_2|}&\propto e^{-x},\qquad
\frac{|V_3^{\mathrm{CL}}|}{|V_2|}\propto\kappa^{-1}(x)\approx x,
\end{aligned}
\label{eq:v3}
\end{equation}
where the release-leakage amplitude is the fraction of energy still unreleased after the first release window and the charge-leakage amplitude is the fraction diverted during charging (see Methods). Because one magnitude decays exponentially with $x$ while the other grows, and their phase difference is in general far from anti-phase, the total leakage is bounded from below, with its nonzero minimum near the crossing of the two (Fig.~\ref{fig:ceiling}b). As long as the phase difference is not driven to anti-phase, the ceiling remains in place. Its origin is the charge leakage alone: lengthening the storage suppresses the release leakage exponentially but feeds the charge leakage further, and the natural switch isolation, Eq.~(\ref{eq:NS}), is accordingly the bound of the charge-leakage channel, estimated where that channel dominates (Methods).

During the storage in the resonator, however, the release leakage advances at $\omega_d$ whereas the charge leakage advances at $\omega$. The charge leakage moreover carries a fixed capacitive phase from its off-state switch crossing: the transfer function of the $C_{\mathrm{off}}$--$Z_s$ divider imprints a quarter cycle, $\phi_C\approx\pi/2$, in the operating regime (see Methods). Their phase difference,
\begin{equation}
\Delta\phi=(\omega-\omega_d)\,\tau_1+\phi_C,
\label{eq:dphi}
\end{equation}
accumulates over a single $\tau_1$-duration time window on top of this quarter-cycle offset. Cancellation requires the two contributions to be opposite in phase and equal in amplitude\cite{slobodkinMassivelyDegenerateCoherent2022,qinTopologicalHystereticWinding2025}, and we now locate where on the $(f,x)$ plane both conditions are met.
 
Mapped over the plane, the two amplitudes occupy complementary regions: the release leakage dominates at small $x$, where a large fraction of the stored energy remains unreleased before leaking to port~3, and the charge leakage at large $x$ (Fig.~\ref{fig:ceiling}c,d). This $x$-dependence originates in the time
domain, reflecting the time-Floquet nature of the leakages\cite{rudnerBandStructureEngineering2020,zhangFloquetTopologicalPhysics2025}, and is independent of the carrier frequency, so an equal-amplitude crossover exists at every $f$. The phase difference of Eq.~(\ref{eq:dphi}), in turn, spans the full $\pm\pi$ range over the plane and reaches anti-phase along the destructive-interference lines
\begin{equation}
f=f_d+\bigl(1.5M-0.75\bigr)\,f_m, \qquad M=\pm1,
\label{eq:lines}
\end{equation}
with $f_d=\omega_d/2\pi$. The capacitive quarter cycle breaks the symmetry of the pair about the resonance: the upper line, $M=1$, $f=f_d+0.75\,f_m$, is drawn toward $f_d$, while its lower partner, $M=-1$, $f=f_d-2.25\,f_m$, is pushed away (Fig.~\ref{fig:ceiling}e). We focus on the upper line, which lies closest to the resonance and carries the higher transmission. On this line, the equal-amplitude condition of Eq.~(\ref{eq:v3}) is met at a single point, its unique root $x_{\mathrm{ISO}}$ (see Supplementary Information), at which the two leakages perfectly cancel (Fig.~\ref{fig:ceiling}f).

\begin{figure*}[t]
  \centering
  \includegraphics[width=0.8\textwidth]{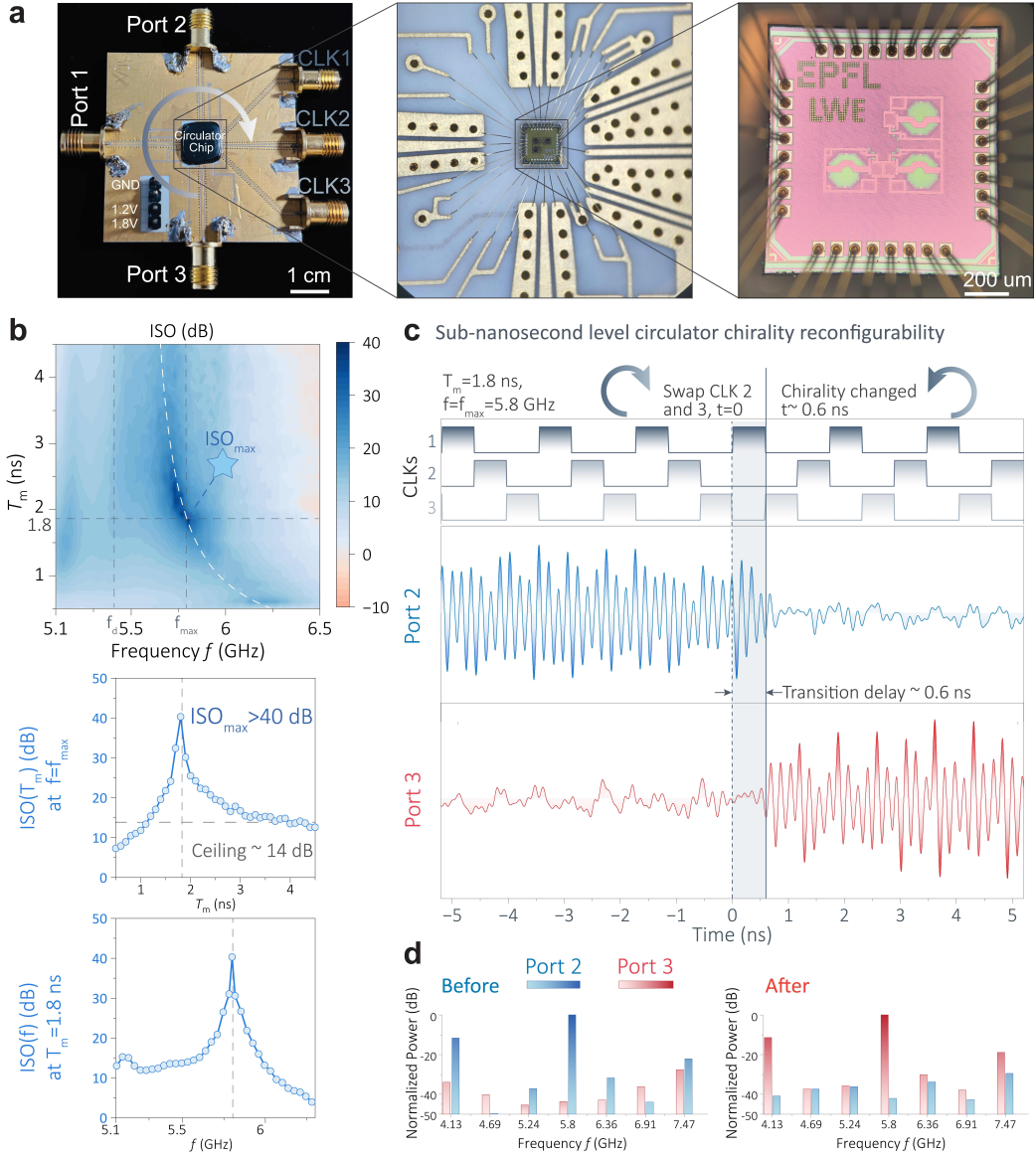}
  \caption{\textbf{Observation of time-Floquet leakage interference and its inherently fast dynamics in a 65-nm CMOS microwave circulator.} \textbf{a}, The Floquet circulator chip on its test board with three clock inputs (CLK1--3) and ports~1--3 (left, with the arrow marking the $1\!\to\!2\!\to\!3$ circulation), the wire-bonded package (middle) and the die micrograph (right). \textbf{b}, The measured isolation follows the upper destructive-interference line and reaches $40$\,dB. Isolation over the carrier frequency $f$--modulation period $T_m$ plane, concentrating along the upper interference line $f=f_d+0.75f_m$ (white dashed), with $f_d=5.37$\,GHz for this prototype. The lower line lies below the measured band. The cuts (bottom), $\mathrm{ISO}(T_m)$ at $f_{\mathrm{max}}$ and $\mathrm{ISO}(f)$ at $T_m=1.8$\,ns, peak above $40$\,dB at the cancellation point ($f_{\mathrm{max}}=5.8$\,GHz, $T_m=1.8$\,ns), approximately $26$\,dB above the $\sim$14-dB natural switch isolation (grey dashed) and limited by the measurement noise floor. \textbf{c}, The interference re-establishes within a single modulation period: swapping two control clocks
  (CLK2$\leftrightarrow$CLK3) at $t=0$, at the cancellation point itself, reverses the circulation chirality in $0.6$\,ns. \textbf{d}, The $40$-dB isolation holds on both sides of the reversal. Port spectra over the Floquet harmonics $f+nf_m$ before (left) and after (right) the swap: the isolated-port carrier lies more than $40$\,dB below the transmitted-port carrier.}
  \label{fig:exp}
\end{figure*}

We then confirm the perfect leakage cancellation on the full isolation map, computed in both theory and transient simulation: the deep isolation concentrates along the destructive-interference lines and peaks at a single point on each (Fig.~\ref{fig:peak}a). On the lines, the anti-phase condition alone yields locally maximal yet finite isolation. Complete cancellation occurs at the crossing with the equal-amplitude point $x_{\mathrm{ISO}}$. For a given resonator this crossing is unique on each line, and the two conditions act independently: the amplitude condition
fixes $x_{\mathrm{ISO}}$, the phase condition fixes the frequency detuning from the
resonance, and the carrier frequency follows as
$f_{\mathrm{max}}=f_d+0.75f_m=f_d+1/(4\tau x_{\mathrm{ISO}})$. The point is
reached by tuning the modulation period alone, which sets both $x=x_{\mathrm{ISO}}$ and, through $f_m$, the carrier frequency
$f_{\mathrm{max}}$. The cut at $f_{\mathrm{max}}$ quantifies the depth of the cancellation
(Fig.~\ref{fig:peak}b). At $x_{\mathrm{ISO}}$ the theoretical isolation
diverges, and the transient simulation resolves an isolation above $40$\,dB, approximately $30$\,dB beyond the $\sim$10-dB ceiling. Away from $x_{\mathrm{ISO}}$ the isolation returns to the ceiling, confirming that the excess arises from the interference. Remarkably, the cancellation condition does not involve the ratio
$f_{\mathrm{coff}}/f$ of the natural switch isolation, Eq.~(\ref{eq:NS}). The equal-amplitude point still exists for a larger
$C_{\mathrm{off}}$, which only shifts $x_{\mathrm{ISO}}$, and the phase condition involves only the frequency detuning between the carrier and the resonance, on top of the fixed quarter-cycle offset.

\section*{Observation in a 65-nm CMOS Floquet circulator}

To experimentally validate the time-Floquet leakage cancellation, we implemented the circulator in a 65-nm CMOS chip with a footprint of $1$\,mm$^2$, integrating three switched resonators of ring-down frequency $f_d=5.37$\,GHz together with their transistor switches. The chip is wire-bonded onto a microwave test board and driven by three external control
clocks, whose period $T_m$ is tuned by an arbitrary waveform generator (Fig.~\ref{fig:exp}a). 
Although the circuit parameters of the prototype differ from those of the simulated design, the physics carries over: the
measured isolation rises along the same upper interference line
$f=f_d+0.75\,f_m$, confirming that the interference involves only the
frequency detuning from the resonance (Fig.~\ref{fig:exp}b).  The measured peak falls at the predicted crossing of this line with the equal-amplitude point $x=x_{\mathrm{ISO}}$, at $(f_{\mathrm{max}},T_m)=(5.8\,\mathrm{GHz},1.8\,\mathrm{ns})$. Both cuts through this point, $\mathrm{ISO}(T_m)$ and $\mathrm{ISO}(f)$, resolve an isolation above $40$\,dB, approximately $26$\,dB above the $\sim$14-dB natural switch isolation of the prototype (Fig.~\ref{fig:exp}b).

Strikingly, the time-Floquet leakage interference needs less than a single
modulation period to settle and is thus inherently fast: the cancellation pattern is rebuilt within every period, so the nonreciprocity it produces can be reprogrammed at the same rate. Swapping two control clocks (CLK2$\leftrightarrow$CLK3) at $t=0$, at the
cancellation point itself ($T_m=1.8$\,ns, $f=f_{\mathrm{max}}$), exchanges
the transmitted and the isolated ports, and the circulation chirality
reverses in $0.6$\,ns, within a single modulation period
(Fig.~\ref{fig:exp}c). The reversal therefore probes the rebuild dynamics of the interference at full null depth: the
deep cancellation, not merely the routing, is re-established within one
period. The port spectra before and after the swap confirm this over the
Floquet harmonics $f+nf_m$: the carrier dominates the transmitted port
while the isolated-port carrier lies more than $40$\,dB lower, on either
side of the reversal (Fig.~\ref{fig:exp}d). Both the depth of the null and the speed of its rebuild are thus set by the clocks, not by the switch.

\section*{Conclusions}

We have beaten the natural switch isolation of integrated Floquet circulators, the intrinsic switch limit proportional to $f_{\mathrm{coff}}/f$. The mechanism is time-Floquet leakage interference: the two coherent leakages of a switched-resonator circulator annihilate each other at the isolated port. The charge leakage through the off-state switch has stood as the ceiling of integrated Floquet circulators. In the switched resonator, the same leakage instead interferes destructively with the release leakage, removing the ceiling it set. In a 65-nm CMOS prototype, the interference lifts the isolation to $40$\,dB, approximately $26$\,dB above the natural switch isolation, and rebuilds within a single modulation period as the clocks reverse the circulation chirality. Because the clocks run far below the carrier frequency, this reprogramming can be carried out by a digital multiplexer that reassigns the clock phases in situ\cite{patilHeterogeneousIntegration015mm2025}. As the cancellation imposes no carrier limit of its own, the isolation limit moves from the switch to the resonator: ultra-high isolation at a chosen carrier frequency requires only a resonator placed near it and a modulation period tuned to the cancellation point. The operation follows wherever an integrated resonator can be built, opening millimetre-wave circulators that fully exploit high-cutoff-frequency switches\cite{samizadehnikooElectronicMetadevicesTerahertz2023,samizadehnikooHighpowerMillimetrewaveSwitches2026}. The interference moreover brings inherently fast, clock-programmable dynamics, a reconfigurability unavailable to ferrite circulators\cite{bowrothu35GHzBariumHexaferrite2020}, locked to one handedness by their magnetic bias. Time-Floquet leakage interference thus offers integrated front ends a nonreciprocity both deep and programmable, beating the isolation limit at the high frequencies where it once tightened fastest. \newpage

\section*{Methods}

\subsection*{N-path time-Floquet architecture and analysis framework}
This section fixes the architecture and notation, derives the storage timescales $(\tau,\omega_d)$, and establishes the window accounting and the energy--spectral factorization from which the amplitudes of the main text follow.

\textbf{N-path time-Floquet architecture.} The circulator of Fig.~\ref{fig:concept}a
is built from $N=3$ identical switched-resonator paths\cite{franksAlternativeApproachRealization1960,soerUnifiedFrequencyDomainAnalysis2010,ghaffariTunableHighQNPath2011,reiskarimianMagneticfreeNonreciprocityBased2016,naguluNonreciprocalElectronicsBased2020}. The switches
are the only time-varying elements: driven by its gate control
voltage, each presents a modulated conductance $G_n(t)$ that toggles
between the on-state $G_{\mathrm{on}}=1/R_{\mathrm{on}}$ and the
off-state, where only the capacitance $C_{\mathrm{off}}$ remains,
while the rest of the network is static, linear, and reciprocal.
Chiral routing rests on holding the wave between charge and release,
an instantaneous path being reciprocal, and the element that holds it
is the reservoir of each path.
The control clocks open the switches in a fixed cyclic order:
$G_1(t)$ is a $T_m$-periodic window of width $\tau_1=T_m/3$, and the
remaining couplings follow by the cyclic shift
\begin{equation}
G_n(t)=G_1\bigl[t-(n-1)\tau_1\bigr],
\label{eq:Gn}
\end{equation}
which fixes the circulation sense $1\!\to\!2\!\to\!3$. Three
time-interleaved paths repeat the same schedule shifted by $\tau_1$,
tiling one period so that exactly one path couples each port at any
instant with constant aggregate coupling. Throughout the Methods the windows are contiguous (no hold interval\cite{naguluChipscaleFloquetTopological2022}) and, owing to the cyclic symmetry, a single source $V_s(t)=\sin(\omega t)$ at port~1 suffices.
The analysis below follows path~1, whose period divides into the three
windows
\begin{equation}
\sigma_1=[0,\tau_1),\quad
\sigma_2=[\tau_1,2\tau_1),\quad
\sigma_3=[2\tau_1,3\tau_1),
\label{eq:windows}
\end{equation}
the charge window and the release windows toward ports~2 and~3,
respectively.

\textbf{Charge--release dynamics.} Within one period the reservoir
traverses two kinds of dynamics. During the charge window $\sigma_1$
it is driven by the source through the on-state switch. During the
release windows the source is disconnected and the stored state
evolves freely as the natural response of the loaded reservoir, an
underdamped ringing $v(t)\propto e^{-t/\tau}\cos(\omega_d t+\varphi)$
characterized by two parameters, the decay time $\tau$ and the ring-down
frequency $\omega_d$. For the resonator of Fig.~\ref{fig:concept}a, with
inductance $L$, capacitance $C$, and loss resistance $R$, loaded by
$Z_t\equiv Z_s+R_{\mathrm{on}}$, they read
\begin{equation}
\begin{aligned}
&\frac{1}{\tau}=\frac{1}{2}\!\left(\frac{1}{CZ_t}+\frac{R}{L}\right),
\qquad
\omega_d=\sqrt{\omega_R^{2}-1/\tau^{2}},\\
&\omega_R^{2}=\frac{1+R/Z_t}{LC},
\end{aligned}
\label{eq:wd}
\end{equation}
with $\omega_R$ the load-corrected resonance, distinct from the bare LC
frequency $\omega_0=1/\sqrt{LC}$. Here $\tau$ is the charge--discharge time
constant of the main text, and the pair $(\tau,\omega_d)$ defines the loaded quality factor $Q_{\mathrm{load}}\equiv\omega_d\tau/2$.

In practical switched operation, inter-cycle self-interference must be
suppressed, so that signals injected in one modulation period do not
corrupt the next as spurious self-mixing or time-domain echoes.
Combined with the exponentially decaying release, this sets the
operating regime of the analysis: the residual energy left in the
reservoir at the end of one period is negligible, and each path is
analysed over a single period starting from a rest initial
condition\cite{stromAnalysisPeriodicallySwitched1977,xuGeneralizedModelLinearPeriodicallyTimeVariant2019}.
Under the cyclic symmetry the $N$ paths differ only by the time shift
$\tau_1$: in the frequency domain a shifted path contributes the same
spectrum up to an overall phase factor, and the paths add identically
at each port. Analysing one path therefore suffices (see Supplementary
Information for the formal reduction and the time-shift accounting that
maps the path-1 leakage onto port~3). For that single path (path~1 in the following), one period suffices. Within one period the dynamics is solved exactly by the piecewise state equations, and the solution carries no Floquet boundary condition. In the operating regime above, however, it transfers the same port energies per period as the periodically modulated path, and the harmonic content of the output follows from a separate spectral accounting, derived below.

\textbf{Energy--spectral factorization.} The port energies are computed in the time domain, from the reservoir node voltage $v(t)$ of the one-period piecewise solution from the rest initial condition (closed form and the approximation used in the theory below). The energy delivered to port $n$ through a window $\sigma$ is
\begin{equation}
E_n=g\int_{\sigma} v^{2}(t)\,dt,
\label{eq:En}
\end{equation}
where $g$ is the equivalent load conductance of the branch through
which the window delivers energy to the port: the on-state branch
presents $g=Z_s/Z_t^{2}$, and the off-state branch
presents the leakage conductance $g=G_{\mathrm{off}}$ derived in the next section. The incident energy is
$E_{\mathrm{inc}}=\int_{\sigma_1}a_1^{2}(t)/Z_s\,dt$, with $a_1=\tfrac12\,[v_1(t)+Z_s i_1(t)]=V_s/2$ the incident wave at port~1, $v_1(t)$ and $i_1(t)$ being the port-1 terminal voltage and current, giving the sideband-summed transmissions
$|S_{n1}(\omega)|^{2}=E_n/E_{\mathrm{inc}}$.
The window accounting follows the routing of Fig.~\ref{fig:concept}. Over one period
each port is coupled to path~1 during exactly one window, so the
integrals reduce to the windows of Eq.~(\ref{eq:windows}). For the
transmitted port, $\sigma_2$ carries the chiral transport, and $E_2$
is evaluated there with the on-state $g=Z_s/Z_t^{2}$. For the isolated port the
accounting separates by mechanism. The release leakage part is probed
by the integral over $\sigma_3$ with the on-state $g$, where the
natural-response residue escapes through the on-state switch. During
$\sigma_1$ the off-state capacitance diverts part of the drive
directly to port~3 with $g=G_{\mathrm{off}}$: this is the charge
leakage. The off-state path also leaks during $\sigma_2$, but this contribution is small and does not affect the physics of the two-leakage interference. For port~3, the two leakages of the $\sigma_1$ and $\sigma_3$ windows therefore capture the key physics of the time-Floquet interference. Where several contributions reach the same port, their amplitudes superpose with relative phases. The energies above fix the
magnitudes, and the phases are tracked in the interference section through the two frequencies at which they oscillate.

The harmonic distribution is fully determined by the modulation and by the transfer function of the reservoir: the periodic on-window of width $\tau_1$ in $T_m$ generates harmonics at $\omega+m\omega_m$ with Fourier weights $c_m$, each filtered by $H(j\omega)$, capacitor or resonator alike, evaluated at the corresponding sideband, so that the fraction of the
transferred power carried by harmonic $m$ is
\begin{equation}
F^{(m)}(\omega)\propto
\bigl|c_m\,H\bigl(j(\omega+m\omega_m)\bigr)\bigr|^{2},
\qquad m/N\in\mathbb{Z},
\label{eq:Fm}
\end{equation}
normalized such that $\sum_m F^{(m)}=1$. The selection rule expresses
the coherent summation over the $N$ time-shifted paths, which cancels
all other harmonics. The two sides combine into the Floquet scattering
coefficients
\begin{equation}
\bigl|S^{(m)}_{n1}(\omega)\bigr|^{2}
=F^{(m)}(\omega)\,\bigl|S_{n1}(\omega)\bigr|^{2}.
\label{eq:factorization}
\end{equation}
The amplitudes of Eq.~(\ref{eq:iso}) are the zeroth-harmonic components. The factorization applies to the release-path contributions, which traverse the modulated transfer: the allocation of Eq.~(\ref{eq:factorization}) gives the transmitted amplitude $|V_2|=\sqrt{F^{(0)}E_2 Z_s}$ and the release-leakage amplitude $|V_3^{\mathrm{RL}}|=\sqrt{F^{(0)}E_3^{\mathrm{RL}} Z_s}$, with $E_3^{\mathrm{RL}}$ the integral of Eq.~(\ref{eq:En}) over $\sigma_3$ and a common window normalization cancelling in every ratio. The factorization does not apply to the charge leakage: it bypasses the modulated transfer in the reduced leakage model (the off-state capacitance is present in every switch state and the capacitor itself performs no frequency translation), so it enters directly as $|V_3^{\mathrm{CL}}|=\sqrt{E_3^{\mathrm{CL}} Z_s}$, without the $F^{(0)}$ weighting. The full
periodically switched network can still generate sidebands through the
time-dependent terminal voltages, and these are retained in the
transient simulation.

\textbf{Single-path solution.} Taking the resonator as the reservoir,
the charge-window voltage is the exact sum of a driven steady state
and a damped transient\cite{phillipsSignalsSystemsTransforms1995},
\begin{equation}
v(t)=\underbrace{\Im\bigl[H(j\omega)\,e^{j\omega t}\bigr]}_{v_{\mathrm{ss}}}
+\underbrace{e^{-t/\tau}\bigl(K_1\cos\omega_d t+K_2\sin\omega_d
t\bigr)}_{v_{\mathrm{tr}}},
\label{eq:vsplit}
\end{equation}
where the steady state follows directly from $H(j\omega)$ and the
transient coefficients $K_{1,2}$ are fixed analytically by the rest
initial condition $v(0)=\dot v(0)=0$. For the capacitor the same
split holds with $H$ a first-order low-pass response and the transient
a non-oscillatory decay. Since the sole role of the transient is to
raise the response from rest toward its steady state, the sum is
captured by a build-up factor,
\begin{equation}
v(t)\;\approx\;\bigl(1-e^{-t/\tau}\bigr)\,v_{\mathrm{ss}}(t),
\qquad 0\le t\le\tau_1,
\label{eq:buildup}
\end{equation}
whose residual oscillatory error averages out in the window integrals
(details in Supplementary Information). During the
release windows $v(t)$ rings at $\omega_d$ with envelope
$e^{-(t-\tau_1)/\tau}$, inheriting its amplitude and phase from the charge-window endpoint by continuity of the reservoir state. In particular, the release envelope decays as $e^{-t/\tau}$ and the two release windows share the same on-state conductance, so the $\sigma_3$ energy integral is $e^{-2x}$ times its $\sigma_2$ counterpart, $E_3^{\mathrm{RL}}=e^{-2x}E_2$, and the release-leakage amplitude of Eq.~(\ref{eq:v3}) follows as $|V_3^{\mathrm{RL}}|/|V_2|=e^{-x}$, the $F^{(0)}$ allocation cancelling in the ratio. The growing charge-leakage amplitude, $|V_3^{\mathrm{CL}}|/|V_2|\propto\kappa^{-1}(x)$, follows in the next section. Every amplitude used in the main text thus follows from this single voltage history: the transmission of Eq.~(\ref{eq:v2}), the two leakage amplitudes of Eq.~(\ref{eq:v3}), and the ceiling of the next section.

\subsection*{Natural switch isolation ceiling}

In this section, we derive the natural switch isolation ceiling of the integrated transistor switches of Fig.~\ref{fig:concept}. For an integrated circuit platform, the switch cutoff frequency $f_{\mathrm{coff}}$ is the figure of merit\cite{dincMillimeterWaveNonMagneticPassive2017,naguluNonMagneticNonReciprocalMicrowave2021}, defined as
\begin{equation}
f_{\mathrm{coff}} \equiv \frac{1}{2\pi R_{\mathrm{on}} C_{\mathrm{off}}}.
\label{eq:fcoff}
\end{equation}
The product $R_{\mathrm{on}}C_{\mathrm{off}}$ is co-limited by the material and transport properties of the platform and by geometry- and
layout-induced parasitics. At the device level, increasing the total switch width (either by increasing the finger count or the per-finger width) reduces $R_{\mathrm{on}}$ approximately inversely with total width, while increasing $C_{\mathrm{off}}$ approximately linearly with width. Consequently, $R_{\mathrm{on}}C_{\mathrm{off}}$ (hence
$f_{\mathrm{coff}}$) is only weakly improved by sizing alone. Once a foundry process design kit (PDK) is fixed, the attainable $R_{\mathrm{on}}C_{\mathrm{off}}$ envelope is
largely determined, and useful switching performance is achieved only at a non-negligible safety margin below $f_{\mathrm{coff}}$. 

During the charge window $\sigma_1$, the reservoir node voltage $v(t)$ couples to each nominally isolated port through the series branch formed by $C_{\mathrm{off}}$ and the port impedance $Z_s$. This mechanism involves only the node voltage and the off-state branch, and therefore holds for any reservoir described by the state equation, capacitor or resonator alike. The branch admittance is
\begin{equation}
Y_{\mathrm{off}}(\omega)
=\frac{1}{Z_s+1/(j\omega C_{\mathrm{off}})}
=G_{\mathrm{off}}(\omega)+j\omega C_{\mathrm{off,eq}}(\omega).
\label{eq:Yoff}
\end{equation}
We focus on the real part that delivers power to the load, with the condition $\omega C_{\mathrm{off}}Z_s\ll 1$ (well satisfied by integrated switches at microwave carriers),
\begin{equation}
G_{\mathrm{off}}(\omega)
=\frac{Z_s}{Z_s^{2}+\left[1/(\omega C_{\mathrm{off}})\right]^{2}}
\approx Z_s\,(\omega C_{\mathrm{off}})^{2}.
\label{eq:Goff}
\end{equation}
The charge-leakage energy per charge window is therefore the time-domain integral
\begin{equation}
E_3^{\mathrm{CL}}
= G_{\mathrm{off}}(\omega)\int_{\sigma_1} v^{2}(t)\,dt,
\label{eq:E3CL}
\end{equation}
which grows with the window duration $\tau_1=x\tau$ once $v(t)$ has built up to its driven steady state (see Supplementary Information for the closed form). This growth is the time-domain origin of the rising charge-leakage amplitude in Fig.~\ref{fig:ceiling}b. As established in the factorization, this direct path enters the zeroth-harmonic isolated-port amplitude without the $F^{(0)}$ allocation, $|V_3^{\mathrm{CL}}|=\sqrt{E_3^{\mathrm{CL}} Z_s}$, and carries the propagation phase $\phi=\omega t$ offset by the capacitive phase $\phi_C$ of the same divider, derived in the phase-evolution section, both entering Eq.~(\ref{eq:dphi}).

To obtain the isolation, we now analyse the transmitted reference, built in strict parallel with
Eq.~(\ref{eq:E3CL}): during the release window $\sigma_2$
the same node voltage $v(t)$ drives port~2 through the on-state branch, delivering
\begin{equation}
E_2=\frac{Z_s}{Z_t^{2}}\int_{\sigma_2} v^{2}(t)\,dt,
\label{eq:E2}
\end{equation}
with $Z_s/Z_t^{2}$ the on-state load conductance $g$ of Eq.~(\ref{eq:En}), the
counterpart of $G_{\mathrm{off}}$ in Eq.~(\ref{eq:E3CL}). The useful
reference must, however, be projected onto the same carrier harmonic:
the release energy is distributed over the Floquet harmonics by the
modulated transfer, and only the fraction $F^{(0)}$ of Eq.~(\ref{eq:Fm}) reaches the transmitted carrier amplitude, $|V_2|=\sqrt{F^{(0)}E_2 Z_s}$, while the direct charge leakage carries no such allocation. Under the cyclic symmetry each of the $N$ paths
contributes equally to the transmission and to the charge leakage within
one period $T_m$, so the common factor $N$ cancels in the ratio and the
carrier-frequency isolation follows from the single-path energies,
\begin{equation}
\begin{aligned}
        \mathrm{ISO}_{\mathrm{CL}}
&=\frac{|V_2|}{|V_3^{\mathrm{CL}}|}
=\left[\frac{F^{(0)}E_2}{E_3^{\mathrm{CL}}}\right]^{1/2}\\
&=\sqrt{\frac{Z_s/Z_t^{2}}{G_{\mathrm{off}}(\omega)}}\;\kappa(x)\\
&=\frac{\sqrt{Z_s^{2}+\left[1/(\omega C_{\mathrm{off}})\right]^{2}}}
      {Z_t}\;\kappa(x),
\label{eq:Iso_m_exact}
\end{aligned}
\end{equation}
where the first factor is the amplitude ratio of the off- and on-state voltage dividers seen from the same node, and
\begin{equation}
\kappa(x)\equiv
\left[F^{(0)}(x)\,
\frac{\int_{\tau_1}^{2\tau_1}v^{2}(t)\,dt}
     {\int_{0}^{\tau_1}v^{2}(t)\,dt}\right]^{1/2}
\label{eq:kappa}
\end{equation}
is the carrier-corrected window kernel, with
$F^{(0)}(x)\equiv F^{(0)}(\omega_d,x)$ for the resonance-centred ceiling
estimate. All details of the reservoir enter only through $v(t)$ and the
allocation $F^{(0)}$: a different reservoir changes only the expression
of $\kappa(x)$, not the divider ratio. The kernel is of order unity, $\kappa\approx0.2$--$0.5$ across the
operating range (closed form in Supplementary Information). The natural
switch isolation is the bound of the charge-leakage channel, and is
therefore estimated where that channel dominates: near $x\approx3$ the
released energy has saturated and the release leakage has decayed away,
giving $\kappa\approx0.3$--$0.4$ and the ceiling scale quoted in the
main text. Correspondingly
$|V_3^{\mathrm{CL}}|/|V_2|\propto1/\kappa(x)$ grows approximately
linearly in $x$ over the range plotted in Fig.~\ref{fig:ceiling}, a
finite-range trend rather than the strict large-$x$ asymptote. The
trend has a simple origin: once $v(t)$ has built up, the charge
integral grows linearly with the window duration while the released
energy saturates, and the carrier fraction $F^{(0)}(x)$ falls slowly
over the same range, so a power-law fit of $\kappa(x)$ over the plotted
interval gives an exponent close to one (see Supplementary Information).

Since $f_{\mathrm{coff}}$ depends only on the chosen
technology node, by recasting Eq.~(\ref{eq:Iso_m_exact}) with $ \omega C_{\mathrm{off}}Z_s\ll1$, we reach the universal, platform-referenced form
\begin{equation}
\mathrm{ISO}_{\mathrm{CL}}
\approx\frac{f_{\mathrm{coff}}}{f}\,
\frac{R_{\mathrm{on}}}{Z_s+R_{\mathrm{on}}}\;\kappa.
\label{eq:Iso_m_fcoff}
\end{equation}
This is the natural switch isolation of the main text, Eq.~(\ref{eq:NS}), fixed by the switch through $f_{\mathrm{coff}}/f$ and tightening linearly as the carrier rises, while the circuit details enter only through the normally small prefactor $R_{\mathrm{on}}/Z_t$ and the order-unity $\kappa$. Because every circuit choice enters only through this prefactor, practical integrated switched circulators cluster within $10$--$20$\,dB of isolation.

\subsection*{Switched capacitor versus switched resonator}
The choice of reservoir decides the modulation speed range. A switched
capacitor is first order: its release is a monotonic decay with no
oscillation, so the stored state holds the amplitude of the wave but
freezes its phase. Recovering the carrier then relies on stitching the
sampled segments across the $N$ interleaved paths, an inherently
sampling-based operation whose effective rate is $Nf_m$, subject to
the Nyquist requirement
$Nf_m\ge 2f$\cite{soerUnifiedFrequencyDomainAnalysis2010,naguluUltraWidebandSwitchedCapacitorDelays2021,tymchenkoQuasielectrostaticWavePropagation2019}.
The modulation is thus pinned to the carrier scale, $f_m\gtrsim 2f/N$,
which becomes prohibitive as the carrier rises.
A switched resonator is second order: during release it keeps ringing
at $\omega_d$, retaining the carrier's phase across many cycles when
the carrier sits close to the resonance, $\omega\approx\omega_d$.
The release requirement is that each on-window last on the order of
the time constant, $\tau_1\sim\tau$. Their ratio is the normalized
modulation period $x\equiv\tau_1/\tau$ of the main text, typically
$1$--$4$ in operation. Frequency recovery no longer relies on
sampling: the Floquet sidebands at $\omega+m\omega_m$, spaced by
$f_m\ll f$ away from the resonance, are suppressed by the reservoir
filtering of Eq.~(\ref{eq:Fm}), and the release preferentially
re-emits at the carrier, restoring the input frequency through the
natural response itself. The capacitor's sampling requirement is
thereby relaxed by the loaded quality factor,
\begin{equation}
f_m\sim\frac{2f}{N\,Q_{\mathrm{load}}}.
\label{eq:fm}
\end{equation}
The resonator thus buys two things at once: a modulation frequency far
below the carrier, and a second phase reference $\omega_d$, distinct from the carrier, on which the
leakage interference of the next section is built.

\subsection*{Phase evolution of the two leakages and the interference
condition}
With the two leakage amplitudes fixed by the window accounting above,
cancellation at port~3 requires the two contributions to be anti-phase
and equal in magnitude. We now track the phases through the two frequencies at which they oscillate.
Each contribution enters Eq.~(\ref{eq:iso}) through its projection
onto the zeroth-harmonic kernel $e^{-j\omega t}$.

\textbf{Capacitive phase of the crossing.} During the charge window $\sigma_1$ the node voltage is driven at the carrier,
$v(t)\propto e^{j\omega t}$, so the product
$e^{j\omega t}e^{-j\omega t}=1$ leaves no net storage phase. The crossing
itself, however, is not phase-free: the charge leakage reaches port~3 through the $C_{\mathrm{off}}$--$Z_s$ divider of the ceiling section, of transfer function
\begin{equation}
\begin{aligned}
H_C(j\omega)&=\frac{j\omega C_{\mathrm{off}}Z_s}
{1+j\omega C_{\mathrm{off}}Z_s},\\
\phi_C=\arg H_C&=\arctan\!\frac{1}{\omega C_{\mathrm{off}}Z_s},
\end{aligned}
\label{eq:phiC}
\end{equation}
so it carries the propagation reference offset by the capacitive phase
$\phi_C$. This phase is controlled by the electrical corner
$1/(2\pi C_{\mathrm{off}}Z_s)$ of the divider, not by the switch figure of
merit $f_{\mathrm{coff}}$. In the operating regime
$\omega C_{\mathrm{off}}Z_s\ll1$ it saturates at a quarter cycle,
$\phi_C\to\pi/2$, deviating by only a few degrees across the band
($\phi_C=84.9^\circ$ at $f_d$ for the simulated design). The main text and the phase map of Fig.~\ref{fig:ceiling}e quote the
quarter-cycle value $\phi_C\approx\pi/2$, while the quantitative theory
of Fig.~\ref{fig:peak} uses the exact frequency-dependent $\phi_C(f)$.

\textbf{Storage phase and the total phase difference.} During $\sigma_2$ the ring-down starts with the phase $e^{j\omega\tau_1}$ imprinted at
switch-off, which is absorbed exactly by the kernel evaluated at the
window boundary. The transmitted amplitude at port~2 therefore shares
the same phase reference, and the phase of the resonator response
$H(j\omega)$, common to both, cancels in the ratio. The release
leakage reaching port~3, however, rings down through $\sigma_3$, one
additional window of duration $\tau_1$ beyond the port-2 release. Over
this window the oscillation advances at $\omega_d$ while the kernel
advances at $\omega$, and the only surviving storage phase factor is
$e^{j(\omega_d-\omega)\tau_1}$: relative to the charge leakage, the
release leakage accumulates $(\omega-\omega_d)\tau_1$, which together
with the capacitive phase of the crossing gives the total phase
difference of Eq.~(\ref{eq:dphi}). The overall sign of $\Delta\phi$
remains a convention immaterial to the anti-phase condition, but the
relative sign between the storage term and $\phi_C$ is physical: it
selects which line of each pair is drawn toward the resonance, as
observed. No other phase survives the projection, so the tunable part of
$\Delta\phi$ accumulates over exactly one window, at a rate set only by
the frequency detuning of the carrier from the resonance.
At the resonance the storage term vanishes and $\Delta\phi\approx\phi_C$:
the two leakages stand in near quadrature and combine essentially in
root sum of squares, the untuned baseline of Fig.~\ref{fig:ceiling}b,
with no accidental cancellation.

\textbf{Destructive-interference lines.} Destructive interference requires $\Delta\phi=M\pi$ with $M$ odd,
while even $M$ is constructive. Writing $\tau_1=T_m/N$ and
$f_m=1/T_m$, the anti-phase condition defines the family of lines
\begin{equation}
\begin{aligned}
f&=f_d+N f_m\!\left(\frac{M}{2}-\frac{\phi_C}{2\pi}\right)\\
&\;\xrightarrow{\;N=3,\ \phi_C\to\pi/2\;}\;
f=f_d+\bigl(1.5M-0.75\bigr)f_m ,
\end{aligned}
\label{eq:linesM}
\end{equation}
whose closest pair, $M=\pm1$, is Eq.~(\ref{eq:lines}) of the main text. The capacitive quarter cycle displaces the
family: the pair closest to the resonance is asymmetric, $M=1$ at
$f_d+0.75f_m$ and $M=-1$ at $f_d-2.25f_m$, the upper line drawn in and
the lower pushed out. 

\textbf{Equal-amplitude point and perfect cancellation.} On each line the phase condition holds for every $x$, so
cancellation is decided by the magnitudes alone. The two amplitudes
vary oppositely with $x$: the release leakage $e^{-x}$ falls
monotonically while the charge leakage grows monotonically, so their
ratio crosses unity at exactly one point $x_{\mathrm{ISO}}$ on each
line. An equal-amplitude crossing therefore exists and is unique, and
a perfect cancellation point necessarily emerges once the modulation
period is tuned to $x=x_{\mathrm{ISO}}$. The phase condition then
fixes the operating carrier through
$f_{\mathrm{max}}=f_d+0.75f_m=f_d+1/(4\tau x_{\mathrm{ISO}})$, reached
by the modulation period alone. The cancellation condition involves the
off-state capacitance only through the saturated quarter-cycle phase, and
never through the ratio $f_{\mathrm{coff}}/f$: a larger $C_{\mathrm{off}}$
shifts $x_{\mathrm{ISO}}$ and nudges $\phi_C$ by a few degrees, while the
anti-phase condition remains set by the carrier--resonance detuning.

\textbf{The capacitor case.} The same projection shows why the capacitor cannot exploit this cancellation. Its second phase reference is frozen at zero frequency, so the phase difference accumulates at the full carrier rate,
$\Delta\phi=\omega\tau_1+\phi_C$, and reaches anti-phase already within a
quarter carrier period of storage. Exploiting it under the sampled operation therefore pins the modulation to the carrier scale, recovering the Nyquist bound of the reservoir comparison above, while the resonator moves the accumulation onto the detuning $\omega-\omega_d$ and reaches anti-phase within a single storage window at $f_m\ll f$.

\subsection*{Transient simulation and simulation parameters}
In this section, we elucidate the circuit model underlying Figs.~\ref{fig:ceiling} and \ref{fig:peak} and its time-domain simulation.

\textbf{Transient simulation.} The numerical results of Figs.~\ref{fig:ceiling} and \ref{fig:peak} are obtained from a transient time-domain simulation of the $N$-path switched-resonator network that makes no approximation beyond the circuit model itself and therefore serves as an independent check of the closed-form theory. The circulator is described as a linear time-varying state-space system\cite{pavanReciprocityInterReciprocityTutorial2023b,pavanReciprocityInterReciprocityTutorial2023}, $\dot{\bm{x}}(t)=\bm{A}(t)\bm{x}(t)+\bm{B}(t)\bm{u}(t)$, where the state vector $\bm{x}$ collects the reservoir node voltages and inductor currents of the $N$ paths, $\bm{u}(t)$ is the port drive, and the matrices $\bm{A}(t)$, $\bm{B}(t)$ switch between a finite set of constant values as the clocks toggle each transistor between $R_{\mathrm{on}}$ and $C_{\mathrm{off}}$.

The equations are integrated with the trapezoidal implicit
(Crank--Nicolson) scheme,
\begin{equation}
\begin{aligned}
    \Bigl(\bm{I}-\tfrac{\Delta t}{2}\bm{A}_{k+1}\Bigr)\bm{x}_{k+1}
&=\Bigl(\bm{I}+\tfrac{\Delta t}{2}\bm{A}_{k}\Bigr)\bm{x}_{k} \\
&+\tfrac{\Delta t}{2}\bigl(\bm{B}_{k}\bm{u}_{k}
+\bm{B}_{k+1}\bm{u}_{k+1}\bigr),
\label{eq:CN}
\end{aligned}
\end{equation}
chosen for its stability on lightly damped oscillatory dynamics: unlike explicit schemes, it introduces no spurious numerical energy growth, which is essential for the high-$Q_{\mathrm{load}}$ resonators simulated over many modulation periods. Since $\bm{A}(t)$ takes values from a finite set, the LU factorizations of $\bm{I}-\tfrac{\Delta t}{2}\bm{A}$ are computed once and reused. 

Scattering parameters are extracted after the initial transient has decayed, by projecting the steady-state port voltages onto each Floquet harmonic $\omega+m\omega_m$ over $50$ modulation periods. This projection evaluates the harmonics directly on the Floquet grid with negligible spectral leakage. The isolation follows from the zeroth-harmonic amplitudes as in Eq.~(\ref{eq:iso}).

\textbf{Simulation parameters.} The simulations of Figs.~\ref{fig:ceiling} and \ref{fig:peak} use a representative integrated switch, $R_{\mathrm{on}}=3\,\Omega$ and $C_{\mathrm{off}}=43.05$\,fF, i.e.\ $f_{\mathrm{coff}}=1.23$\,THz, coupled to on-chip-scale resonators ($L=110.5$\,pH, $C=5.07$\,pF, resonator quality factor $Q=40$) under $Z_s=50\,\Omega$. The differential release modes of the complete frozen-clock network give $f_d=6.57$\,GHz and $\tau=0.433$\,ns (bare LC frequency $f_0=6.73$\,GHz), corresponding to $Q_{\mathrm{load}}=\omega_d\tau/2\approx9$. Evaluating Eq.~(\ref{eq:Iso_m_fcoff}) near the charge-dominated point $x\approx3$, with the zeroth-harmonic fraction included in $\kappa$, gives $\kappa\approx0.3$--$0.4$ and a natural switch isolation of ${\sim}10$\,dB. The complete physical-$C_{\mathrm{off}}$ transient network gives the quantitative saturation level of approximately $9$\,dB in Fig.~\ref{fig:ceiling}a, the remaining difference being the expected accuracy of the single-path scaling estimate.

\subsection*{Chip implementation}
The parameters of the fabricated chip differ from the simulated ones, but probe the same leakage cancellation physics at a second operating point. The transistor-level design of the chip itself was
separately verified by schematic and post-layout simulations (Cadence Spectre, foundry PDK models, parasitics extracted with Calibre PEX). Each of the three paths integrates an on-chip LC resonator together with its three transistor switches. Including the layout and packaging parasitics, the damped ring-down frequency of the effective resonator is $f_d=5.37\,\mathrm{GHz}$, extracted directly from
the measured isolation map of Fig.~\ref{fig:exp}b. Three packaged prototypes were measured and all reproduce the interference line and the isolation peak. The data shown are from one representative device.

For the switches of the fabricated chip, the packaged prototype is
described by the equivalent switch parameters
$R_{\mathrm{on}}=9.5\,\Omega$ and $f_{\mathrm{coff}}=455$\,GHz
($C_{\mathrm{off}}=36.8$\,fF), the datasheet cutoff frequency of
$0.5\,\mathrm{THz}$ being effectively reduced by the pad and packaging
capacitances loading the off state. Evaluating
Eq.~(\ref{eq:Iso_m_fcoff}) at the operating carrier $f=5.8\,\mathrm{GHz}$
with $\kappa\approx0.4$ gives a ceiling of ${\approx}14\,\mathrm{dB}$,
consistent with the measured isolation floor of Fig.~\ref{fig:exp}b.

\subsection*{Measurement setup}
The setup comprises a four-port vector network analyzer (VNA, Rohde \& Schwarz ZNA67, 10~MHz--67~GHz), a four-channel arbitrary waveform generator (AWG, Keysight M8199A, 128~GSa/s), a four-channel real-time oscilloscope (Keysight Infiniium DSOV084A, 80~GSa/s), and a four-channel mixed-signal oscilloscope (Tektronix MSO64). Three output channels of the AWG generate the periodic clock signals that drive the on-chip switches through the microwave test board. The clocks follow the cyclic sequence with a tunable modulation period $T_m$, adjacent channels being delayed by $T_m/3$, corresponding to a relative phase shift of $120^\circ$. Each clock channel passes through a bias tee (Mini-Circuits ZX85-12G-S+), which adds a DC offset to the clock waveform. The channel gain and the DC bias are adjusted such that the high level of each clock at the switch terminal is 1.2~V. The MSO64 monitors the three clock waveforms and verifies their voltage levels, relative delays, and synchronization. The nonreciprocal isolation is measured with the VNA connected to the three ports of the circulator.

For the time-resolved measurement of the chirality reversal, a microwave signal generator (Rohde \& Schwarz SMR20) provides a continuous-wave single-tone excitation at port~1. The transmitted signals at ports~2 and~3 are captured simultaneously by the DSOV084A, while the AWG generates the clock sequence that reverses the circulation direction. After correction for the calibrated cable delays, the measured transients give the intrinsic temporal response of the circulator, from which the chirality-reversal time is quantified. The reversal time is defined as the interval between the clock swap and the instant at which the port-2 and port-3 envelopes cross their post-reversal steady-state levels.

The fourth output channel of the AWG generates a trigger for the DSOV084A, providing a common temporal reference for the transient measurements. One input channel of the DSOV084A records a representative clock signal together with the transmitted signals at ports~2 and~3, enabling direct temporal alignment between the applied modulation and the observed chirality reversal. The broadband scattering parameters of all cables connecting the test board to the instruments are measured in advance, and their frequency-dependent phase responses and group delays are used to calibrate the timing offset between the trigger and the actual reversal arriving at the circulator. The delays on the test board are extracted from full-wave simulation in ANSYS HFSS. This calibration accounts for the relative propagation delays of the clock, trigger, excitation, and measurement paths.

\vspace{6pt}
\noindent\textbf{Acknowledgements.}~This work was supported by the Swiss National Science Foundation under the Eccellenza grant 181232, the Swiss State Secretariat for Education, Research and Innovation (SERI) under contract No. MB22.00028, and Huawei Technology Sweden. The authors also acknowledge funding from the SwissChips initiative. Z.Z. acknowledges the Swiss National Science Foundation fellowship (P500-2\_235518). Z.Z. and R.F. thank Dr.\ Alexandre Levisse (EPFL EDA Center) for his assistance with the Cadence simulations and chip layout, and acknowledge Europractice IC Service for its support. Z.Z. and R.F. acknowledge the helpful discussions with Dr.\ Mattias Gustafsson. Z.Z. thanks Ji Zhou and Gang Wang at EPFL for their help with the experiments.

\vspace{2pt}
\noindent\textbf{Data availability.}~The data supporting the findings of this study
are presented within the article.

\vspace{2pt}
\noindent\textbf{Code availability.}~The code used to evaluate the conclusions in the
paper is available from the corresponding authors upon request.
\vspace{2pt}

\noindent\textbf{Author contributions.} Z.Z. and R.F. conceived the idea. Z.Z., H.Q., and J.W. conducted the theory. Z.Z. and R.F. designed the chips with help from the EPFL EDA Center. Z.Z. designed the test board and measured the chips with H.Q., J.W., and Q.C.. Z.Z. drafted the manuscript. R.F. reviewed and revised the manuscript. All authors contributed to discussions of the results under the supervision of R.F.

\vspace{2pt}
\noindent\textbf{Competing interests.}~The authors declare no competing interests.

\vspace{2pt}
\noindent\textbf{Correspondence.}~Correspondence and requests for materials should be
addressed to R.F.\ (\href{mailto:romain.fleury@epfl.ch}{romain.fleury@epfl.ch}).
\newpage
\bibliographystyle{apsrev4-2}
\bibliography{refs_paper}

\end{document}